\documentclass[aps,showpacs,superscriptadress,preprint]{revtex4}
\usepackage{graphicx}
\usepackage{makecell}

\begin{document}
\title{Scalar field quasinormal modes of noncommutative high dimensional Schwarzschild-Tangherlini  black hole spacetime with smeared matter sources
%%with general smeared matter distribution%%
}
\author{Zening Yan$^{1}$, Chen Wu$^{2}$\footnote{Electronic address: wuchenoffd@gmail.com} and Wenjun Guo$^{1}$} \affiliation{
\small 1. University of Shanghai for Science and Technology, Shanghai 200093, China\\
\small 2. Shanghai Advanced Research Institute, Chinese Academy of Sciences, Shanghai 201210, China}

% z.n.yan.bhtpr@gmail.com, impgwj@126.com
\begin{abstract}
 We investigate the massless scalar quasinormal modes (QNMs) of the noncommutative $D$-dimensional Schwarzschild-Tangherlini  black hole spacetime  in this paper.
 By using the Wentzel-Kramers-Brillouin (WKB) approximation method, the asymptotic iterative method (AIM) and the inverted potential method (IPM) method, we  made a detail analysis of  the massless scalar QNM frequencies by varying  the general smeared matter distribution and the allowable characteristic parameters ($k$ and $\theta$) corresponding to different dimensions. It is found that the nonconvergence of the high order WKB approximation exists in the QNMs frequencies of scalar perturbation around the noncommutative $D$-dimensional Schwarzschild black holes.  We conclude that the 3rd WKB result  should be more reliable than those of the high order WKB method since our numerical results are also verified by the AIM method and the IPM method. In the dimensional range of $4\leq D \leq7$, the scalar QNMs as a function of the different papameters (the noncommutative parameter $\theta$, the smeared matter distribution parameter $k$, the multipole number $l$ and the main node number $n$) are obtained.
Moreover, we study the dynamical evolution of  a scalar field in the background of the noncommutative high dimensional Schwarzschild-Tangherlini  black hole.
\end{abstract}

 \maketitle

\section{Introduction}
Motivated by string theory arguments \cite{Snyder:1946qz}, noncommutative spacetime has been studied extensively and is considered to be an alternative way to a quantum gravity. Quantum mechanics tells us that the emergence of  a minimal length is a natural requirement when we consider quantum features of phase space. The basic idea of the noncommutative geometry is that the singularities in black holes at the origin can be avoided by the presence of spacetime
minimal length: the point-like object on the classical commutative manifold should then be replaced by a smeared object \cite{Smailagic:2003yb, Smailagic:2003rp}.
An important application of the noncommutative geometry is the black hole spacetime. In 1993, Susskind proposed that stringy effect should not be ignored in the string/black hole correspondence principle \cite{Susskind:1993ki}. It is known that noncommutative geometry inspired black holes contain stringy effects, where such an effect is similar in some sense to that of noncommutative field theory induced by string theory \cite{Nicolini:2005vd}. Thereafter, string theory, noncommutative geometry and black holes have been studied more deeply.

In 1999, Seiberg and Witten published important work and proposed that some low-energy effective theory of open strings with a nontrivial background can be described by a noncommutative filed theory, so that string theory provides a theoretical background for the study of noncommutative spacetime \cite{Seiberg:1999vs}. According to the results of string theory, the spacetime coordinates become noncommutative operators on $D$ - brane \cite{Witten:1995im}. The noncommutativity of spacetime can be encoded in the commutator
\begin{equation}
\left[\hat{x}_{\mu}, \hat{x}_{\nu}\right]=i \theta_{\mu \nu},\label{non operators}
\end{equation}
where $\hat{x}_{\mu}$ are the spacetime coordinates operators and $\theta_{\mu \nu}$ is an antisymmetric constant tensor of dimension $(length)^{2}$.

The noncommutative correction of quantum field theory can be made by Eq. (\ref{non operators}). Currently the method to noncommutative quantum field theory follows two distinct paths: one is based on the Weyl-Wigner-Moyal $\star$-product \cite{Smailagic:2003yb} and the other on coordinate coherent state formalism \cite{Smailagic:2003rp}. Nicolini and his coworkers pointed out that the noncommutative effect is the inherent property of the manifold itself, not the superposition of the geometric structure. Therefore, the noncommutative effect only acts on the matter source term and there is no need to change the Einstein tensor part of the field equation \cite{Nicolini:2005vd}. Specifically,  the mass density of the point-like function on the right side of the Einstein equation is replaced by the Gaussian smeared matter distribution, but the left side does not change. Of course, in addition to Gaussian smeared matter distribution, it also includes Lorentz smeared matter distribution, Rayleigh distribution and ring-type distribution, etc. In this way, the solution of a self-regular black hole with noncommutative effect without curvature singularity is given.

And then, Nicolini and his coworkers first give the solution of four-dimensional noncommutative-geometry-inspired Schwarzschild black hole \cite{Nicolini:2005vd}. And then extended to the case of charged \cite{Ansoldi:2006vg} and cosmological constants \cite{Nicolini:2011dp}. In 2010,  Modesto and Nicolini extended it to the general case of charged rotating noncommutative black holes \cite{Modesto:2010rv}. Noncommutative black holes are also extended to high-dimensional \cite{Nozari:2009nr, Rizzo:2006zb} black holes and high-dimensional charged black holes \cite{Spallucci:2009zz, Nozari:2007ya}.
Nozari and Mehdipour have studied the Hawking radiation of noncommutative Schwarzschild black holes \cite{Nozari:2008rc}.
Many authors have also studied the thermodynamic properties of noncommutative black holes \cite{Banerjee:2008gc, Myung:2006mz, Nozari:2006rt, Miao:2015npc}.
Since the noncommutative spacetime coordinates introduce a new fundamental natural length scale, it is also of interest to study the influence of this constant on QNMs frequencies in this black hole spacetime.

The investigations concerning the interaction of black holes with various fields
around  give us the possibility of obtaining some information about the physics of black holes. One of these information could be obtained from  QNMs which are
characteristic of the background black hole spacetimes. QNMs is defined as the complex solution of perturbed wave equation under certain boundary conditions, it dominates the decay of the field in the background of the black hole.
The survey of field perturbation in black hole spacetimes motivated the extensive numerical and analytical study of QNMs \cite{Berti:2003ud, Berti:2004um, Konoplya:2004wg, Konoplya:2007yy}.
 Since QNMs only depends on the macroscopic properties of black holes, astrophysicists are interested in the analysis direction of gravitational waves by QNMs.
They believe that in the gravitational wave signals from perturbed black holes, QNMs dominate an exponentially decaying ringdown phase \cite{Kokkotas:1999bd}, it also dominates the ringdown phase of the gravitational system created by the merger of a pair of black holes \cite{Pretorius:2005gq, Campanelli:2005dd}.
On September 14th, 2015, two detectors of the Laser Interferometer Gravitational-wave Observatory (LIGO) made the first direct measurement of gravitational waves \cite{TheLIGOScientific:2016agk}.  The Advanced LIGO detectors observed a transient gravitational-wave signal determined to be the coalescence of two black holes, thereby launching the era of gravitational wave astronomy. The gravitational wave signal captured by this event can also be used to obtain the bounds on the quantum fuzziness scale of noncommutative spacetime \cite{Kobakhidze:2016cqh}. Furthermore, it is believed that QNMs is closely related to the Ads/CFT correspondence \cite{Maldacena:1997re, Pramanik:2015eka} in string theory and loop quantum gravity \cite{Kunstatter:2002pj, Natario:2004jd}. All these motivated the  extensive numerical and analytical studies of QNMs for different spacetime and fields around black holes \cite{Brito:2014nja, Hod:2015hza, Wu:2015fwa, Wu:2018xza}.

The rest of the paper is organized as follows. In section 2, we introduce brief description of the noncommutative $D$-dimensional Schwarzschild-Tangherlini spacetime black holes with general smeared matter distribution, and give  the perturbative equation of the  scalar perturbation in given background  and the effective potential equation. We also discuss the allowable range of $k$ under the hoop conjecture and the range of $\theta$ corresponding to the different values of $D$ and $k$ when the event boundary exists. In section 3, we use the WKB, the AIM and the IPM method to calculate QNMs and compare these numerical results. In section 4, the dynamical evolution of massless scalar field in the background of noncommutative $D$-dimensional Schwarzschild black hole is analyzed by time-domain integral method. Finally, a brief summary of the full text is presented.

\section{The Basic equations}
\subsection{The metric}
From a physical perspective, noncommutative geometry is described as a fluid that spreads around the origin, rather than a fluid that is squeezed at the origin, and the energy-momentum tensor modified by the smeared matter distribution corresponds to the anisotropic fluid.
Recently, several researchers \cite{Nicolini:2011fy, Miao:2016ipk, Miao:2016ulg, Park:2008ud} have shown that the Gaussian smeared matter distribution can be replaced as long as there is a sharp peak similar to the Dirac $\delta$ function at the origin, and the integral of the distribution function is limited.
In Ref. \cite{Park:2008ud}, the authors proposed a general smeared matter density
\begin{equation}\label{rho}
\rho_{\mathrm {matter}}(r)=\left[\frac{M}{\pi^{\frac{D-1}{2}}(2 \sqrt{\theta})^{D+k-1}} \cdot \frac{\Gamma\left(\frac{D-1}{2}\right)}{\Gamma\left(\frac{D+k-1}{2}\right)}\right] r^{k} e^{-\frac{r^{2}}{4 \theta}},
\end{equation}
where $M$ is the mass of black holes, $\theta$ is a positive noncommutative parameter. $k$ is a non-negative integer, the Gaussian smeared matter distribution corresponds to $k=0$, Rayleigh distribution \cite{Myung:2008kp} to $k=1$, and Maxwell-Boltzmann distribution \cite{Park:2008ud} to $k=2$, etc.
%In addition, there are the Lorentzian distribution \cite{Nozari:2008rc}, the ring-type distribution \cite{Park:2008ud}.

By using Eq. (\ref{rho}), it can be deduced that the mean radius of the matter corresponding to the mass density distribution is
\begin{equation}\label{radiusmean}
\bar{r}=\int_{0}^{\infty}  \frac{ r \rho_{\mathrm {matter}}(r)}{M} d V_{D-1}=2 \sqrt{\theta} \frac{\Gamma\left(\frac{D+k}{2}\right)}{\Gamma\left(\frac{D+k-1}{2}\right)},
\end{equation}
where $dV_{D-1}$ is an $(D-1)$-dimensional volume element.
The more dispersed the matter distribution is, the larger the corresponding parameter $\theta$ is, while the more concentrated the matter distribution is, the smaller the corresponding parameter $\theta$ is. In the case of $\theta \rightarrow 0$ limit, the matter mean radius $\bar{r} \rightarrow 0$, which means that the distribution of matter collapses into a point and the noncommutation of spacetime disappears.
Therefore, it can be found that this noncommutation of spacetime is only a small effect superimposed on ordinary spacetime, and the noncommutation effect is mainly reflected in the area near the mean radius of matter.

For the geometry metric solution of static spherically symmetric noncommutative Schwarzschild black holes, the following conditions must be satisfied: 1)
the radial matter distribution function $\rho_{\mathrm {matter}}(r)$ is spherically symmetric; 2) the covariant energy conservation $\nabla_{\mu} T^{\mu \nu}=0$; 3) the Schwarzschild-like property $g_{00}=-g_{r r}^{-1}$ is preserved.
Therefore, the spherically symmetric energy-momentum tensor of the matter source distribution  \cite{Nozari:2007ya, Wu:2018cfn} can be written as $T_{\mu \nu}=\left(\rho+p_{\vartheta_{i}}\right) u_{\mu} u_{\nu}+p_{\vartheta_{i}} g_{\mu \nu}+\left(p_{r}-p_{\vartheta_{i}}\right) \chi_{\mu} \chi_{\nu}$,
where $u_{\mu}$ is the velocity of the fluid, and $\chi_{\mu}$ is an unit vector along the radial direction, namely
\begin{equation}
\Big [T_{\phantom{12}\nu}^{\mu}\Big ]_{\mathrm {matter}}=\mathrm{diag}\Big [-\rho_{\mathrm {matter}}(r), p_{r}, p_{\vartheta_{1}}, \cdots, p_{\vartheta_{D-2}}\Big ],
\end{equation}
where $p_{r}=-\rho_{\mathrm{matter}}(r)$ is the radial pressure and the tangential pressure $p_{\vartheta_{i}}=-\rho_{\mathrm {matter}}(r)-\Big [r /(D-2)\Big ] \partial_{r} \rho_{\mathrm {matter}}(r)$.
%Such sources are usually considered anisotropic fluids, by solving the Einstein equation
The Einstein field equation can be written as
\begin{equation}
R_{\mu \nu}-\frac{1}{2} R g_{\mu \nu}=8 \pi \mathcal{G} \Big [T_{\mu \nu}\Big ]_{\mathrm {matter}},
\end{equation}
where $\mathcal{G}$ is gravitational constant in $D$-dimensional spacetime, $\mathcal{G}=m_{P l}^{-2}=\ell_{P l}^{-2}$ where $m_{P l}$ and $\ell_{P l}$ are the Planck mass and the Planck length, respectively.
%$[T_{\mu \nu}]_{\mathrm {matter}}$ is the spherical symmetry energy-momentum tensor that describes the $D$-dimensional.

The metric of noncommutative spherically symmetric black hole in $D$-dimensional Schwarzschild-Tangherlini spacetime is
\begin{equation}
d s^{2}=-f(r) d t^{2}+f(r)^{-1} d r^{2}+r^{2} d \Omega_{D-2}^{2},
\end{equation}
where $d \Omega_{D-2}^{2}$ is the line element on the $(D-2)$-dimensional unit sphere is given by
\begin{equation}
d \Omega_{D-2}^{2}=d \vartheta_{1}^{2}+\sin ^{2} \vartheta_{1} d \vartheta_{2}^{2}+\cdots+(\sin ^{2} \vartheta_{1} \cdots \sin ^{2} \vartheta_{D-3}) d \vartheta_{D-2}^{2}=\sum_{j=1}^{D-2}\left(\prod_{i=1}^{j-1} \sin ^{2} \vartheta_{i}\right) d \vartheta_{j}^{2},
\end{equation}
here use the coordinate system $\left(t, r, \vartheta_{1}, \vartheta_{2}, \cdots, \vartheta_{D-2}\right)$, and the lapse function \cite{Miao:2015npc} is
\begin{equation}
f(r)=1-\frac{16 \pi \mathcal{G} m(r)}{(D-2) \Omega_{D-2} r^{D-3}},
\end{equation}
where $\Omega_{D-2}$ is the volume of the $(D-2)$-dimensional unit sphere is given by \cite{Cardoso:2002pa, Konoplya:2003ii}
\begin{equation}
\Omega_{D-2}=\frac{2 \pi^{\frac{D-1}{2}}}{\Gamma\left(\frac{D-1}{2}\right)},
\end{equation}
and $m(r)=\int_{0}^{r} \rho_{\mathrm {matter}}(r) \Omega_{D-2}r^{2} dr$ represents the mass distribution of black holes, which is given by
\begin{equation}
m(r)=\frac{M}{\Gamma\left(\frac{D+k-1}{2}\right)} \gamma\left(\frac{D+k-1}{2},\left(\frac{r}{2 \sqrt{\theta}}\right)^{2}\right),
\end{equation}
this allow us to rewrite the lapse function as
\begin{equation}\label{NDSchfr}
f(r)=\left[1-\frac{16 \pi \frac{M}{\Gamma\left(\frac{D+k-1}{2}\right)}}{(D-2)\left(\frac{2 \pi^{(D-1)/2}}{\Gamma\left(\frac{D-1}{2}\right)}\right)} \cdot \frac{\gamma\left(\frac{D+k-1}{2}, \frac{r^{2}}{4 \theta}\right)}{r^{D-3}}\right],
\end{equation}
$\Gamma(x)$ is the gamma function and $\gamma(a, x)$ is the lower incomplete gamma function. Throughout the rest of this paper, geometric units are used so that $\mathcal{G} = c = \hbar = k_{B} = 1$.

The following introduces several special black hole solutions, which are special forms of the general smeared matter distribution noncommutative $D$-dimensional Schwarzschild black hole solution.
When taking the noncommutative limit $\theta \rightarrow 0$ or $r \gg \theta$, Eq. (\ref{NDSchfr}) approaches the lapse function of the $D$-dimensional Schwarzschild black hole \cite{Tangherlini:1963bw}.
When $k=0$  satisfied, Eq. (\ref{NDSchfr}) represents a noncommutative $D$-dimensional Schwarzschild black hole under Gaussian smeared matter distribution \cite{Rizzo:2006zb}.
Eq. (\ref{NDSchfr})  represents the noncommutative geometry inspired Schwarzschild black hole \cite{Nicolini:2005vd} when the parameters $D=4$, $k=0$.
When the above conditions are met at the same time, that is, $\theta \rightarrow 0$, $D=4$, $k=0$, Eq. (\ref{NDSchfr}) is converted to the Schwarzschild black hole.

\subsection{Perturbation equation and Effective potential for scalar field}
To consider the scalar QNMs frequencies in the afore-mentioned background, it is necessary to consider the Klein-Gordon equation describing the evolution of the massless scalar perturbation field
\begin{equation}
\frac{1}{\sqrt{-g}}\left(\sqrt{-g} g^{\mu \nu} \Psi_{, \nu}\right)_{, \mu}=0,
\end{equation}
where $g$ is the determinant of $g_{\mu \nu}$, $\Psi$ denotes the scalar field in the form of \cite{Natario:2004jd}
\begin{equation}
\Psi\left(t, r,\left\{\vartheta_{i}\right\}\right)=\sum_{l, m} r^{\frac{2-D}{2}} \psi_{l}(t, r) Y_{l, m}\left(\vartheta_{1}, \vartheta_{2}, \cdots, \vartheta_{D-2}\right),
\end{equation}
where the $\left\{\vartheta_{i}\right\}$ represents the $(D - 2)$ dimensional angles and $Y_{l, m}\left(\vartheta_{1}, \vartheta_{2}, \cdots, \vartheta_{D-2}\right)$ is the $D$ - dimensional spherical harmonics and $\psi_{l}(t, r)=e^{i \omega t} \Phi_{\omega}(r)$.

Utilizing a tortoise coordinate change $x=\int \mathrm{d}r/f(r)$ and the separation of radial and angular variables,
%, substituting $\Psi$ into Eq. (\ref{NDSchfr}),
the Klein-Gordon equation can be reduced to the Schr$\ddot{\text{o}}$dinger-like wave function
\begin{equation}\label{waveequ}
\frac{d^{2} \Phi_{\omega}}{d x^{2}}(x)+\omega^{2} \Phi_{\omega}(x)-V(x) \Phi_{\omega}(x)=0,
\end{equation}
where $\omega$ is the complex QNMs frequency, and the effective potential is given by
\begin{equation}\label{DimPotential}
V_{\theta, k}(r)=f(r) \left[\frac{l(l+D-3)}{r^{2}}+\frac{(D-2)(D-4)}{4 r^{2}} f(r)+\frac{(D-2)}{2 r} \frac{\partial f(r)}{\partial r}\right],
\end{equation}
where $l=0,1,2,\cdots$ are the multipole numbers.
The tortoise coordinate $x$ is defined on the interval $(-\infty,+\infty)$ in such a way that the spatial infinity $r = +\infty$ corresponds to $x = +\infty$ and the event horizon $r_{h}$ corresponds to $x = -\infty$.
The above effective potential is positively defined and has the form of the potential barrier which approaches constant value at both spatial infinity and event horizon.
It is found that the maximum  value $V_{0}$ of the  barrier becomes lower as the parameter $k$ or the noncommutative parameter $\theta$ increases.
In other words, the parameter $k$ and the parameter $\theta$ have similar influences on the variation of $V_{\theta, k}(r)$ as a function of $r$ in different dimensions.

 The boundary condition at the horizon is for the solution to be purely ingoing wave, while the other boundary condition at spatial infinity is such that the
  wave has to be purely outgoing one \cite{Ferrari:2007dd}. Therefore one can write the boundary conditions as
%The boundary condition of pure incoming wave corresponding to event horizon and pure outgoing wave QNMs at infinity is
\begin{equation}
\Phi_{\omega}(x) e^{+i \omega x}, \;\;\;\; \text{as}  \;\;\;\; x \rightarrow -\infty,
\end{equation}
\begin{equation}
\Phi_{\omega}(x) e^{-i \omega x},\;\;\;\; \text{as}  \;\;\;\; x \rightarrow +\infty.
\end{equation}
%where $C_{\pm}$ are arbitrary coefficient. Pure incoming waves mean that nothing can escape from the horizon, and pure outgoing waves indicate that there is no radiation coming in from infinity .

\subsection{The allowable value of $k$ and the valid range of $\theta$}
%\subsection{The allowable range of $k$ under hoop conjecture is satisfied}
According to the Eq. (\ref{NDSchfr}), the formulas of $M$ and the event horizon radius $r_{h}$ can be derived
\begin{equation}
M=\frac{(D-2)\left(\frac{2 \pi^{(D-1) / 2}}{\Gamma\left(\frac{D-1}{2}\right)}\right) \Gamma\left(\frac{D+k-1}{2}\right)}{16 \pi \gamma\left(\frac{D+k-1}{2}, \frac{r_{h}^{2}}{4 \theta}\right)} r_{h}^{D-3},
\end{equation}
where $r_{h}$ as the largest root of $f(r)=0$. The event horizon radius $r_{H}$ of the extreme black hole satisfies the relation $\partial M / \partial r_{h}=0$, therefore, $r_{H}$ satisfies the following equation
\begin{equation}
\frac{2 x_{H}^{(D+k-1)} e^{-x_{H}^{2}}}{\gamma\left(\frac{D+k-1}{2}, x_{H}^{2}\right)}=D-3,
\end{equation}
where $x_{H}$ is defined by $x_{H}= r_{H} / 2 \sqrt{\theta}$.

Here we need to consider the hoop conjecture \cite{Thorne:1972ji, Casadio:2013uga}: the matter mean radius of the mass distribution is less than the event horizon radius of extreme black holes, that is, $\bar{r} \leq r_{H}$ or $\bar{x} \leq x_{H}$, where $\bar{x}= \bar{r}/ 2 \sqrt{\theta}$. In addition, from a thermodynamic point of view, the hoop conjecture ensures that the smallest black hole (extreme black hole) has zero temperature and zero heat capacity. If the hoop conjecture is violated, the black hole will not have an extreme configuration. Combined with Eq. (\ref{radiusmean}), is given by
\begin{equation}\label{Dkrelationship}
\frac{\Gamma\left(\frac{D+k}{2}\right)}{\Gamma\left(\frac{D+k-1}{2}\right)} \leq x_{H}(D, k).
\end{equation}
Therefore, the allowable range of the $k$ values of a noncommutative $D$-dimensional Schwarzschild black hole  corresponding to different $D$ values can be derived by using the Eq. (\ref{Dkrelationship}), as shown in Table. \ref{krange}.
It can be seen from this table that the Gaussian smeared matter distribution is only applicable to the noncommutative Schwarzschild-Tangherlini black hole in four and five dimensions.
\begin{table}[tbh]\centering
\caption{Allowable range of $k$ values corresponding to different $D$ values.
\vspace{0.3cm}} \label{krange}
\begin{tabular*}{16.4cm}{*{9}{c @{\extracolsep\fill}}}
\hline
$D$ & 4 & 5 & 6 & 7 & 8 & 9 & 10 & 11 \\
\hline
$k$ & $\geq 0$ & $\geq 0$ & $\geq 4$ & $\geq 8$ & $\geq 14$ & $\geq 22$ & $\geq 32$ & $\geq 43$ \\
\hline
\end{tabular*}
\end{table}

%\subsection{The valid range of $\theta$ when the event boundary exists}
On the other hand, the noncommutative $D$-dimensional Schwarzschild-Tangherlini black hole has multiple event horizons, that is, the different roots of the equation $f(r)=0$. For the convenience of calculation, $M=1$ is set.

In the left panel of Fig. 1 shows  radial dependence of lapse function $f(r)$  for three values of the noncommutative parameter $\theta$ when the dimension number $D$ and the parameter $k$ are fixed.  It is noteworthy that in  the $D$-dimensional Schwarzschild-Tangherlini black hole the lapse function $f(r)$ can have zero, one, or two horizons depending on the value of $\theta$.  Therefore we can obtain   the extreme $\theta$ parameter for which the inner horizon and outer horizon coincide, which we record as $\theta_{\mathrm{max}}$.

% it can be seen that a specific range of $\theta$ corresponds to the existence of event horizon of black holes.
%When $f(r)=0$ has a unique root (event horizon), we can calculate this limit value of $\theta$ and record this value as $\theta_{\mathrm{max}}$.
%Therefore, when $0<\theta\leq\theta_{\mathrm{max}}$ the noncommutative $D$-dimensional
%Schwarzschild black hole have two event horizons, when $\theta=\theta_{\mathrm{max}}$ it has an horizon and when $\theta>\theta_{\mathrm{max}}$ it has no event horizon.

\begin{figure}[htbp]\label{fr}
\centering
\includegraphics[height=6.4cm,width=16cm]{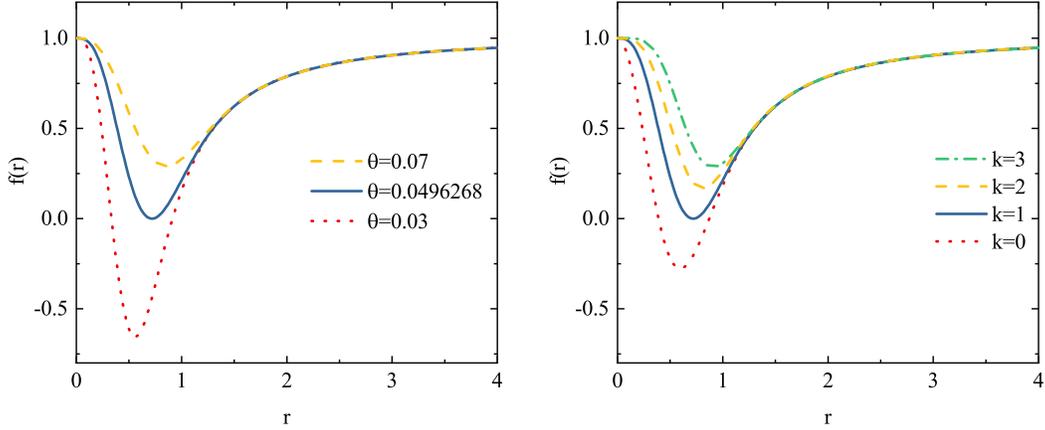}
\caption{The left panel shows  variation of the lapse function $f(r)$ with respect to the radial coordinate $r$ for three values of the  $\theta$ when $M=1$, $D=5$, $k=1$. The right panel shows variation of the lapse function $f(r)$ with respect to the radial coordinate $r$ for four values of the $k$ when $M=1$, $D=5$, $\theta=0.0496268$.}
\end{figure}

In the right panel of Fig. 1 also shows radial dependence of lapse function $f(r)$  for four values of  $k$ parameter in different smeared matter distribution
 when $D$ and $\theta$ are fixed.
%It can be seen that parameter $k$ and noncommutation parameter $\theta$ have similar effects on the changing trend of $f(r)$, respectively.
For the sake of discussion, we calculate the $\theta_{\mathrm{max}}$ values for the five minimum allowable k values in the range of $7 \geq D \geq 4$, as listed in Table. \ref{Thetamax}.
\begin{table}[tbh]\centering
\caption{Values of $\theta_{\mathrm{max}}$ for different $k$ and $D$ values. The parameter is selected as $M=1$.
\vspace{0.3cm}} \label{Thetamax}
\begin{tabular*}{16.4cm}{*{7}{c @{\extracolsep\fill}}}
\hline
\multicolumn{7}{c}{$\theta_{\mathrm{max}}$}\\
\hline
\hline
$D$ & $k=0$ & $k=1$ & $k=2$ & $k=3$ & $k=4$ & $k=5$ \\
\hline
4 & 0.275811 & 0.214662 & 0.178070 & 0.153293 & 0.135218 & 0.121357 \\
5 & 0.0633279 & 0.0496268 & 0.0412099 & 0.035442 & 0.0312102 & 0.0279564  \\
\hline
\hline
$D$ & $k=4$ & $k=5$ & $k=6$ & $k=7$ & $k=8$ & $k=9$ \\
\hline
6 & 0.0231343 & 0.0207527 & 0.0188472 & 0.0172844 & 0.0159774 & 0.0148664 \\
\hline
\hline
$D$ & $k=8$ & $k=9$ & $k=10$ & $k=11$ & $k=12$ & $k=13$ \\
\hline
7 & 0.0150971 & 0.0140589 & 0.0131622 & 0.0123793 & 0.0116894 & 0.0110764 \\
\hline
\end{tabular*}
\end{table}

\section{Numerical methods and numerical results}
\subsection{Brief introduction of the WKB, the AIM and the IPM  method}
The QNM frequencies of the scalar perturbation in the noncommutative high dimensional Schwarzschild-Tangherlini black hole spacetime. The real part of QNMs corresponds to the oscillation frequency, while the imaginary part corresponds to the damping, therefore, the complex  $\omega$ values are written as $\omega = \text{Re}(\omega) + i \text{Im}(\omega)$. One can solve the massless scalar QNMs by several numerical methods, such as integration of the wavelike equations, the monodromy method, the continued fraction method, the WKB approximation method, the asymptotic iteration method  and so on. We calculate the massless scalar QNMs by using the 6rd order WKB method, the IPM method as well as the AIM method.  Next we give a brief introduction to these three methods.

The WKB approximation method was applied for the first time by Schutz and Will \cite{Schutz:1985zz}. Iyer and his coworkers developed the WKB method up to 3rd \cite{Iyer:1986np} order and later, Konoplya developed it up to 6th order \cite{Konoplya:2003ii}. This semianalytic method has been applied extensively in numerous black hole spacetime cases, which has been proved to be accurate up to around one percent for the real and the imaginary parts of the quasinormal frequencies for low-lying modes with $n < l$, where $n$ is the mode number and $l$ is the angular momentum quantum number.

The 6th order formalism  of the WKB approximation has formula
\begin{equation}
\frac{i\left(\omega^{2}-V_{0}\right)}{\sqrt{-2 V_{0}^{\prime \prime}}}-\sum_{i=2}^{6} \Lambda_{k}=n+\frac{1}{2},
\end{equation}
where $V_{0}$ and $V_{0}^{''}$  are the maximum potential and the second derivative of the potential evaluated at the maximum potential, $n$ is the number of nodes and $\Lambda_{k}$ represents the $k$-th order correction. The formulae for $\Lambda_{2}$, $\Lambda_{3}$, $\Lambda_{4}$, $\Lambda_{5}$ and $\Lambda_{6}$ are given in \cite{Iyer:1986np, Konoplya:2003ii}.

%Ref \cite{Cardoso:2003vt} shows that when $l>n$, WKB approximation is a good method, but when $l=n$, WKB method is not very good, and for $l<n$, WKB method is not applicable at all.
%Therefore, we only give the numerical results of WKB approximation under the condition of $l>n$.

In Ref. \cite{Ciftci:2005xn}, the asymptotic iterative method (AIM) was applied to solve second order differential equations for the first time.
This new method was then used to obtain the QNM frequencies of field perturbation in Schwarzschild black hole spacetime \cite{Cho:2009cj}.
%In order to apply this method, we need to consider a special initial value of the massless scalar field and combine its behavior at the event horizon
% and infinity, thus, the homogeneous linear second order differential equation is given by
Let's consider a second order differential equation of the form
\begin{equation}\label{2nd Eq}
\chi''=\lambda_{0}(x)\chi'+s_{0}(x)\chi,
\end{equation}
where $\lambda_{0}(x)$ and $s_{0}(x)$ are well defined functions and sufficiently smooth. Differentiating the equation above with respect to $x$  leads to
\begin{equation}
\chi'''=\lambda_{1}(x)\chi'+s_{1}(x)\chi,
\end{equation}
where the two coefficients are  $\lambda_{1} (x) =\lambda_{0}'+s_{0}+\lambda_{0}^{2}$ and $s_{1}(x)=s_{0}'+s_{0}\lambda_{0}.$
Using this process iteratively, differentiate $n$ times with respect to the independent variable, which produces the following equation
\begin{equation}
\chi^{(n+2)}=\lambda_{n}(x)\chi'+s_{n}(x)\chi,
\end{equation}
where the new coefficients $\lambda_{n}(x)$ and $s_{n}(x)$ are associated with the older ones through the following relation
\begin{equation}
\lambda_{n}(x)=\lambda'_{n-1}(x)+s_{n-1}(x)+\lambda_{0}(x)\lambda_{n-1}(x), \;\;\;\; s_{n}(x)=s'_{n-1}(x)+s_{0}(x)\lambda_{n-1}(x),\label{iteration}
\end{equation}
for sufficiently large values of $n$, the asymptotic concept of the AIM method is introduced by \cite{Cho:2011sf}
\begin{equation}
\frac{s_{n}(x)}{\lambda_{n}(x)}=\frac{s_{n-1}(x)}{\lambda_{n-1}(x)} = \text{Constant}. \label{Quantum condition}
\end{equation}

The perturbation frequency can be obtained from the above-mentioned "quantization condition". Then, $\lambda_{n}(x)$ and $s_{n}(x)$ are expanded into Taylor series around the point $x'$ at which the AIM method is performed
\begin{equation}
\lambda_{n}\left(x^{\prime}\right)=\sum_{i=0}^{\infty} c_{n}^{i}\left(x-x^{\prime}\right)^{i}, \;\;\;\; s_{n}\left(x^{\prime}\right)=\sum_{i=0}^{\infty} d_{n}^{i}\left(x-x^{\prime}\right)^{i},
\end{equation}
where $c_{n}^{i}$ and $d_{n}^{i}$ are the $i$-th Taylor coefficients of $\lambda_{n}(x')$ and $s_{n}(x')$, respectively. Substitution of above equations into Eq. (\ref{iteration}) leads to a set of recursion relations for the Taylor coefficients as
\begin{equation}
c_{n}^{i}=\sum_{k=0}^{i} c_{0}^{k} c_{n-1}^{i-k}+(i+1) c_{n-1}^{i+1}+d_{n-1}^{i}, \;\;\;\; d_{n}^{i}=\sum_{k=0}^{i} d_{0}^{k} c_{n-1}^{i-k}+(i+1) d_{n-1}^{i+1},
\label{recursion}
\end{equation}
after applying the recursive relation (\ref{recursion}) in Eq. (\ref{Quantum condition}) \cite{Ciftci_2003}, the quantization condition can be obtained
\begin{equation}
d_{n}^{0}c_{n-1}^{0}- c_{n}^{0} d_{n-1}^{0}=0,
\end{equation}
which can be used to calculate the QNMs of black holes more accurately.

The inverted potential method (IPM) is the P$\ddot{\text{o}}$schl-Teller potential approximation method \cite{Poschl:1933zz}, which uses the P$\ddot{\text{o}}$schl-Teller potential $V_{P T}$ to approximate the effective potential
$V$ in the tortoise coordinate system
\begin{equation}
V_{P T}=\frac{V_{0}}{\cosh ^{2} \alpha\left(x -x_{0}\right)}, \quad -2 V_{0} \alpha^{2}=\left.\frac{d^{2} V}{d x^{2}}\right|_{x=x_{0}},
\end{equation}
where $V_{0}$ is the height of the effective potential and $-2 V_{0} \alpha^{2}$ is the curvature of the potential at its maximum.
The bound states of the P$\ddot{\text{o}}$schl-Teller potential are well known
\begin{equation}
\Omega=\alpha^{\prime}\left[-\left(n+\frac{1}{2}\right)+\left(\frac{1}{4}+\frac{V_{0}}{\left(\alpha^{\prime}\right)^{2}}\right)^{1 / 2}\right], \quad n=0,1,2,\cdots.
\end{equation}
The quasinormal modes $\omega$ can be obtained from the inverse transformation $\alpha^{\prime}=i \alpha$ as follows
\begin{equation}
\omega=\pm \sqrt{V_{0}-\frac{1}{4} \alpha^{2}}-i \alpha\left(n+\frac{1}{2}\right), \quad n=0,1,2,\cdots.
\end{equation}
%Only when the degree of agreement between the effective potential $V$ and $V_{P T}$ is high, the accuracy of QNMs calculated by this method can be higher.
 It is well known that for the low-lying QNMs, in the majority of cases the behavior of the effective potential is essential only in some region near the black hole, so that the fit of the height of the effective potential and of its second derivative is indeed enough for calculation of QNM frequencies. This method gives quite accurate estimation for the  high multipole number modes.

\subsection{Comparison of numerical results}
Now,  we report the frequencies of the massless scalar perturbation in the noncommutative $D$-dimensional Schwarzschild black hole with general smeared matter distribution by using the afore-mentioned methods.  Results for the scalar QNM frequencies for $D=4$ and $k=0$ (noncommutative geometry inspired schwarzschild black hole) under consideration are listed in Table. \ref{NGISchdata}. It is found that the values of the QNM frequencies  computed by the third order WKB approximation are almost coincide with those results in Ref. \cite{Liang:2018uyk}. We should emphasize that the QNM frequencies computed by the AIM method have  positive and unreasonable imaginary parts of the QNMs when the $\theta$ value is  smaller than one critical value  $\tilde{\theta}$. 
In Table. \ref{NGISchdata}, with respect to $\theta$, the results are different and depend on the method. The AIM results are considered to be precise values. The WKB results in the higher order may be convergent to the AIM method, However, from Table. \ref{NGISchdata}, one can conclude that in the higher order of the WKB method, the difference between this method and the AIM method   become larger. So far, the high-order equations  of the WKB method has not been mathematically proven to converge to the theoretical resolution. For different black hole spacetime, the higher order calculation results given by this method may have certain deviations and errors.
In Ref. \cite{Batic:2019zya}, the WKB results up to the sixth order approximation reveals that this method is not convergent exactly in the calculation of noncommutative geometry inspired schwarzschild black hole, which means that this instabilities  might be originated from the inefficiency  or limitation of the WKB approximation. This case teaches us an important lesson for the determination of the QNMs: a single method is not always appropriate to decode all the QNMs.

\begin{table}[tbh]\centering
\caption{Comparison between the WKB method, the AIM method and the IPM method in the calculation of the scalar QNMs for noncommutative $D$-dimensional Schwarzschild black hole spacetime. The parameter is selected as $M=1$, $D=4$, $k=0$, $l=3$, $n=0$, and the numerical results is accurate to five decimal places.
\vspace{0.3cm}} \label{NGISchdata}
\begin{tabular*}{16.4cm}{*{5}{c @{\extracolsep\fill}}}
\hline
$\theta$ & WKB 6th & WKB 3rd & AIM & IPM \\
\hline
0.02 &	0.67537 $-$ 0.09650i & 0.67521 $-$ 0.09651i & 2.50084 $+$ 2.64382i & 0.67810 $-$ 0.09709i \\
0.04 &	0.67537	$-$ 0.09650i & 0.67521 $-$ 0.09651i & 0.42793 $+$ 0.63196i & 0.67810 $-$ 0.09709i \\
0.06 &	0.67537 $-$ 0.09650i & 0.67521 $-$ 0.09651i & 0.47180 $+$ 0.13028i & 0.67810 $-$ 0.09709i \\
0.08 &	0.67537 $-$ 0.09650i & 0.67521 $-$ 0.09651i & 0.67539 $-$ 0.11222i & 0.67810 $-$ 0.09709i \\
0.10 &	0.67537 $-$ 0.09652i & 0.67521 $-$ 0.09651i & 0.67539 $-$ 0.09672i & 0.67810 $-$ 0.09709i \\
0.12 &	0.67502 $-$ 0.09669i & 0.67521 $-$ 0.09654i & 0.67537 $-$ 0.09653i & 0.67810 $-$ 0.09709i \\
0.14 &	0.67496 $-$ 0.09655i & 0.67518 $-$ 0.09662i & 0.67525 $-$ 0.09654i & 0.67810 $-$ 0.09709i \\
0.16 &	0.67625 $-$ 0.09557i & 0.67506 $-$ 0.09668i & 0.67516 $-$ 0.09637i & 0.67810 $-$ 0.09707i \\
0.18 &	0.67725 $-$ 0.09453i & 0.67474 $-$ 0.09653i & 0.67474 $-$ 0.09607i & 0.67812 $-$ 0.09699i \\
0.20 &	0.67578 $-$ 0.09455i & 0.67418 $-$ 0.09597i & 0.67418 $-$ 0.09569i & 0.67816 $-$ 0.09680i \\
0.22 &	0.67242 $-$ 0.09549i & 0.67341 $-$ 0.09491i & 0.67385 $-$ 0.09503i & 0.67825 $-$ 0.09641i \\
0.24 &	0.66950 $-$ 0.09620i & 0.67252 $-$ 0.09338i & 0.67378 $-$ 0.09390i & 0.67843 $-$ 0.09573i \\
0.26 &	0.66849 $-$ 0.09568i & 0.67161 $-$ 0.09144i & 0.67381 $-$ 0.09271i & 0.67873 $-$ 0.09466i \\
\hline
\end{tabular*}
\end{table}

In this paper, through  presenting the scalar QNMs   for various values of the parameters $D$, $k$ and $\theta$ by means of three different numerical methods, we can address that the IPM method and  the third order WKB method  will not reach a positive $Im(\omega)$ at all.
Since it is well known that the accuracy of the P$\ddot{\text{o}}$schl-Teller potential approximation method is not very high,  we only compare the QNMs results obtained by the WKB method and the AIM method as shown in  Fig. 2.  It can be seen that, the  fluctuation of numerical results given by the 6th, 5th and 4th WKB method increase dramatically with the increase of $\theta$, while the AIM results are more closer to  the three order WKB results for lager $\theta$ parameter.
In most numerical cases, the 6th WKB approximation method is adequate precise. Such abnormal behavior of the 6th WKB  results has recently been discussed in the calculation of the gravitational QNMs  in spacetime with smeared matter sources  \cite{Das:2018fzc, Pramanik:2019qgy}.
We should mention that these works  discuss the noncommutative geometry inspired schwarzschild black hole for the $D=4$ and $k=0$ case. The authors argued that the QNMs results obtained by the WKB approximation method are only valid  for  small $\theta$ parameter interval, while the numerical fluctuation corresponding to a slightly larger $\theta$ is a spurious oscillations. However, neither of these two works considered any other numerical method for  comparison with the WKB method, so one can not know which order WKB method is more accurate.  In order to  test reliability of the high order  WKB method more strictly, we also compute the QNMs frequencies  by  means of the AIM method, as shown in Fig. 2 and Fig. 3. Our results show that the QNMs obtained by the 3rd WKB method and the AIM method are  very close to each other, so we can reach a conclusion that the 3rd WKB approximation is more accurate than other high order approximation.

\begin{figure}[htbp]
\centering
\includegraphics[height=8.25cm,width=16.4cm]{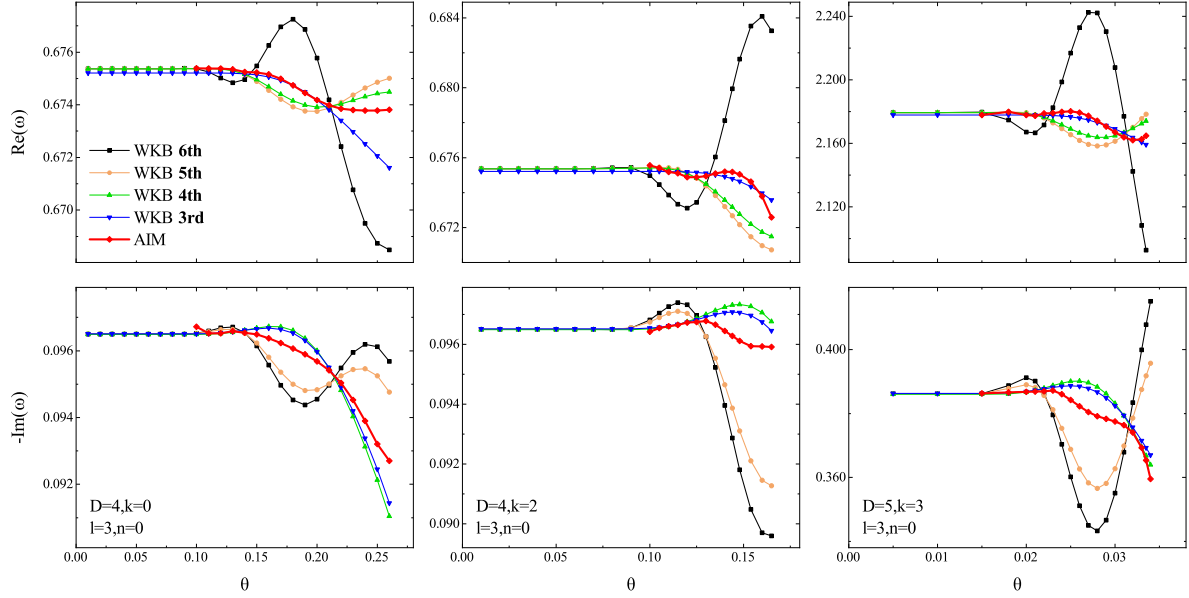}
\caption{$Re(\omega)$ and $-Im(\omega)$ parts of the QNMs of the noncommutative $D$-dimensional Schwarzschild black hole with different $D$ and $k$ as a function of $\theta$ are given for the $n=0$, $l=3$ mode by means of the (6th, 5th, 4th, 3rd) WKB approximations and the AIM method.}
\end{figure}

Moreover, by means of different methods, the QNM frequencies results for different values of $D$ with $l=3$ and $n=0$  are shown in Table. \ref{DSchdata}.
The results show that in the case of high dimension, the QNMs obtained by method WKB and the QNMs obtained by method AIM are closer than those obtained by method IPM.
Therefore, in Fig. 3, we still compare the QNMs obtained by each order WKB method with the QNMs obtained by AIM method, and get the same conclusion in high dimension.
We also show  the QNMs values as a function of  $\theta$  for different values of $k$ in Fig. 4-7.
We can see that the values of $Re(\omega)$ and $-Im(\omega)$ remain same when $\theta$ is small, but when $\theta$ exceeds a critical value $\theta^{\prime}$, the values of $Re(\omega)$ and $-Im(\omega)$ decrease with the increase of $\theta$.
And when the $k$ value increases, the critical value $\theta^{\prime}$ decreases.

\begin{table}[tbh]\centering
\caption{Comparison between the QNMs results by means of different methods. The parameter is selected as $M=1$, $l=3$, $n=0$, $7 \geq D \geq 4$. The results retain six significant digits.
\vspace{0.3cm}} \label{DSchdata}
\begin{tabular*}{16.4cm}{*{5}{c @{\extracolsep\fill}}}
\hline
$D$	& 6th WKB & 3rd WKB & AIM & IPM \\
\hline
$4$ & 0.675366 $-$ 0.0965006i & 0.675206 $-$ 0.0965121i & 0.675366 $-$ 0.0964996i & 0.678098 $-$ 0.0970906i\\
$5$ & 2.17936 $-$ 0.386200i & 2.17786 $-$ 0.386345i & 2.17936 $-$ 0.386189i & 2.19612 $-$ 0.390715i \\
$6$ & 3.29990 $-$ 0.638150i & 3.29601 $-$ 0.638820i & 3.29979 $-$ 0.638276i & 3.33540 $-$ 0.648949i \\
$7$ & 4.12768 $-$ 0.830792i	& 4.12035 $-$ 0.832944i & 4.12683 $-$ 0.831727i & 4.18228 $-$ 0.849524i \\
\hline
\end{tabular*}
\end{table}

\begin{figure}[htbp]
\centering
\includegraphics[height=8.25cm,width=16.4cm]{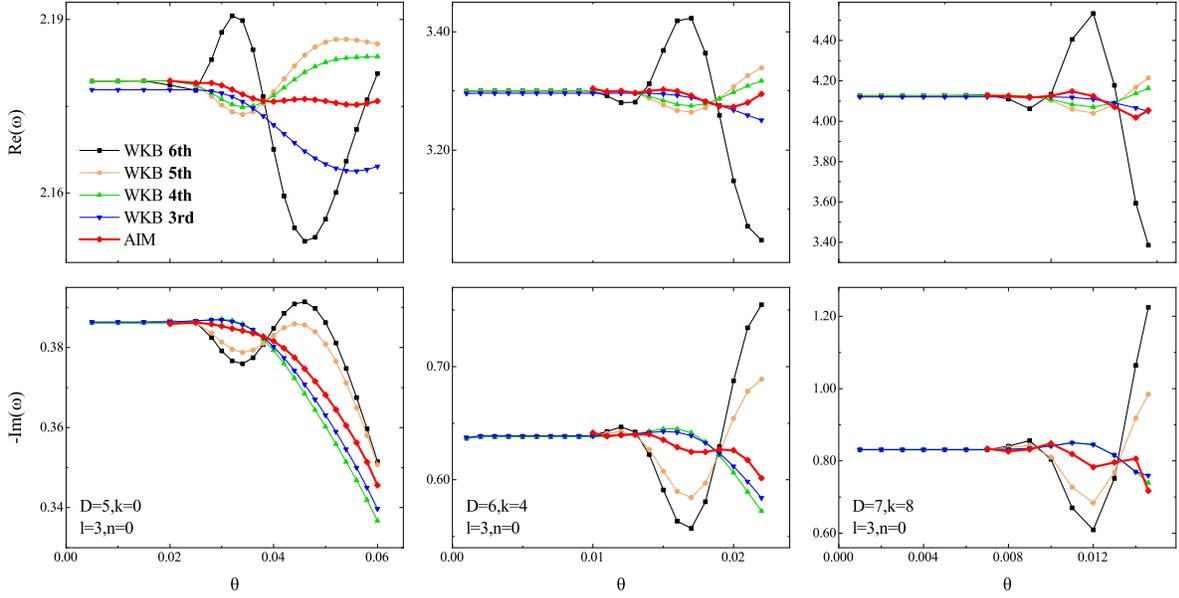}
\caption{Similar to the Fig. 2, but for different $D$ and $k$.}
\end{figure}

\begin{figure}[htbp]
\centering
\includegraphics[height=20.5cm,width=15.6cm]{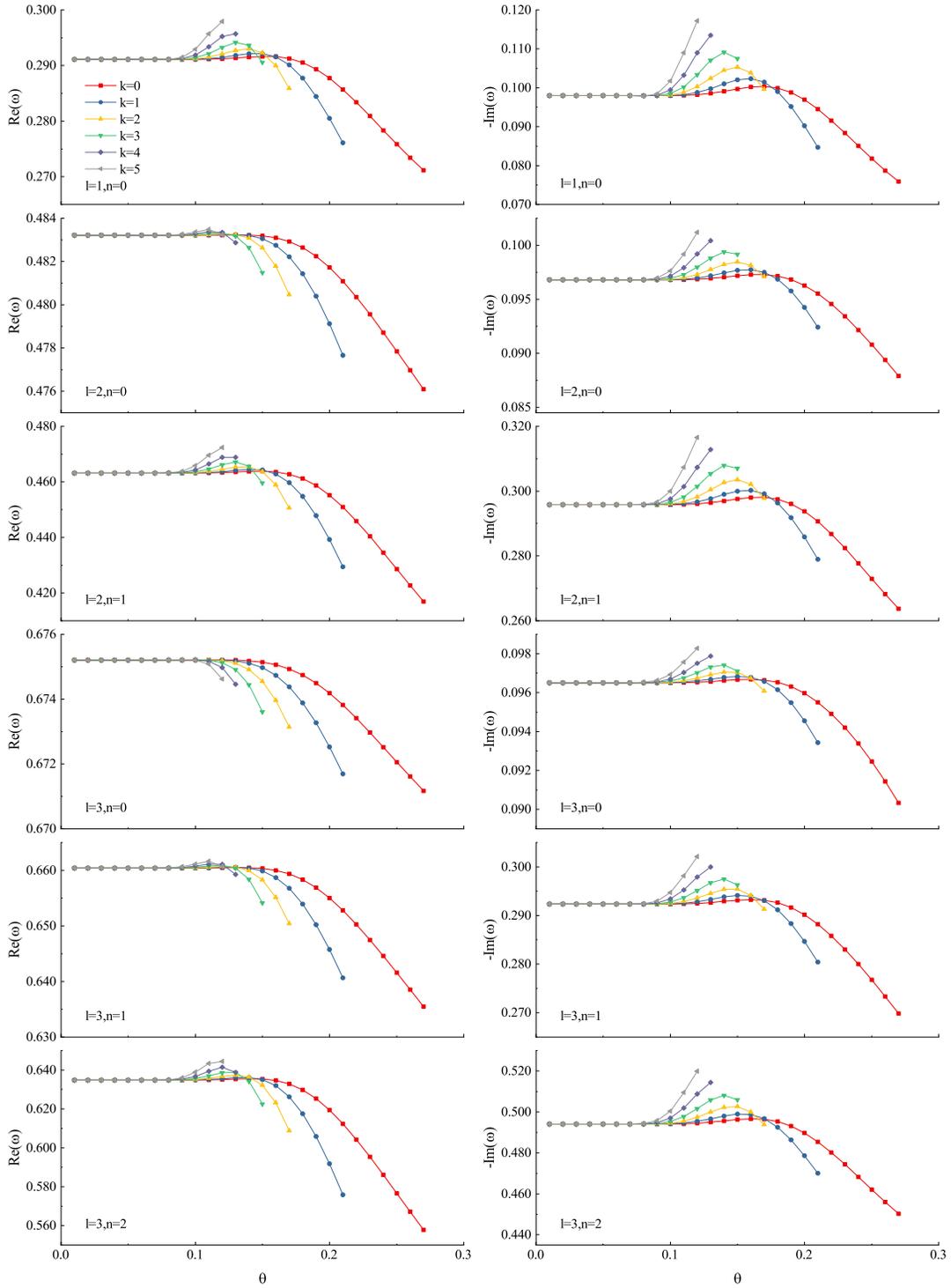}
\caption{$Re(\omega)$ and $-Im(\omega)$ for scalar field in  the noncommutative $D$-dimensional Schwarzschild black hole spacetime  as a function of $\theta$ for $D=4$, $M=1$  with different values of $k=0, 1, 2, 3, 4, 5$. }
\end{figure}

\begin{figure}[htbp]
\centering
\includegraphics[height=20.5cm,width=15.6cm]{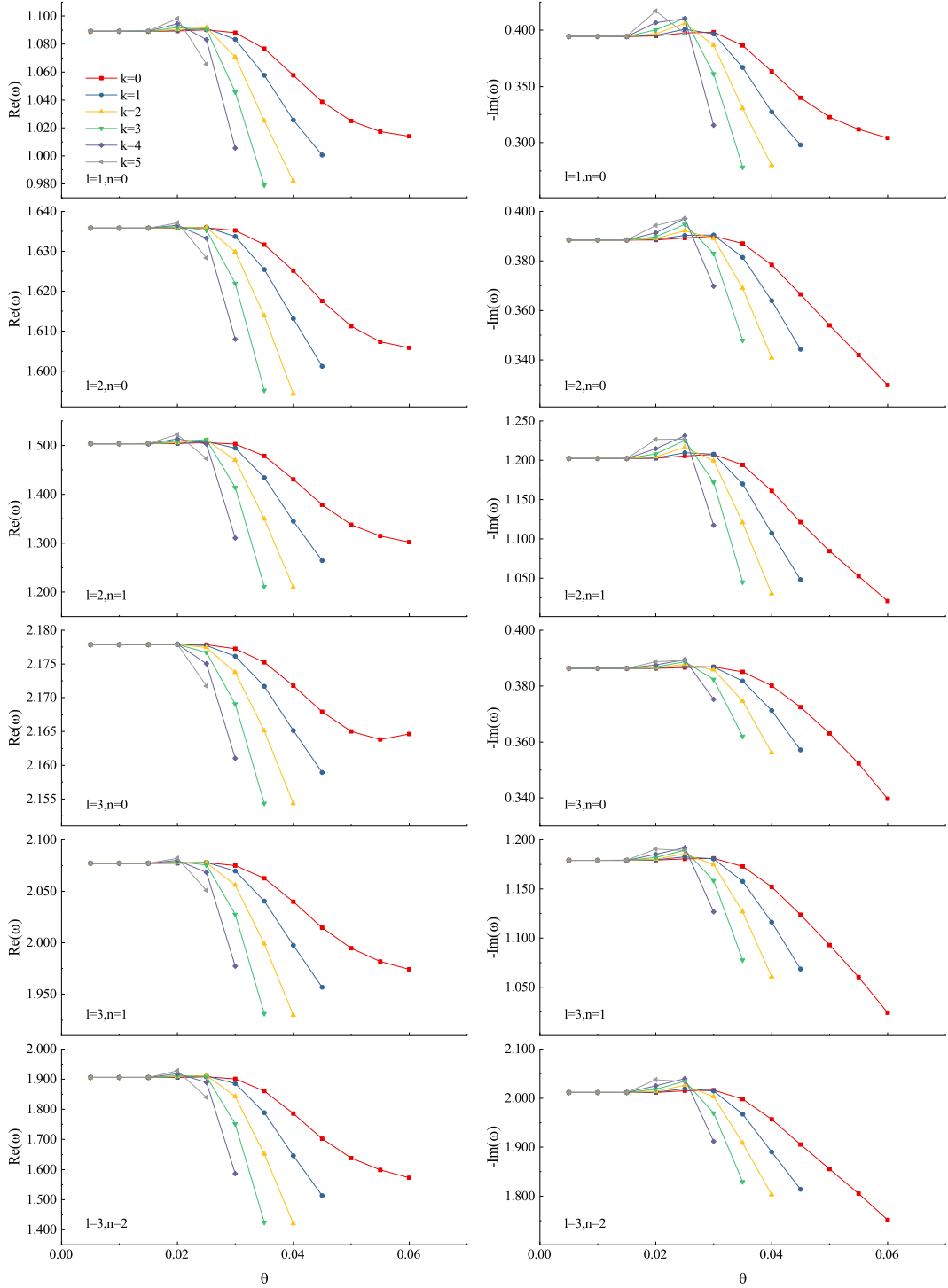}
\caption{Similar to the Fig. 4, but for $D=5$. }
\end{figure}

\begin{figure}[htbp]
\centering
\includegraphics[height=20.5cm,width=15.6cm]{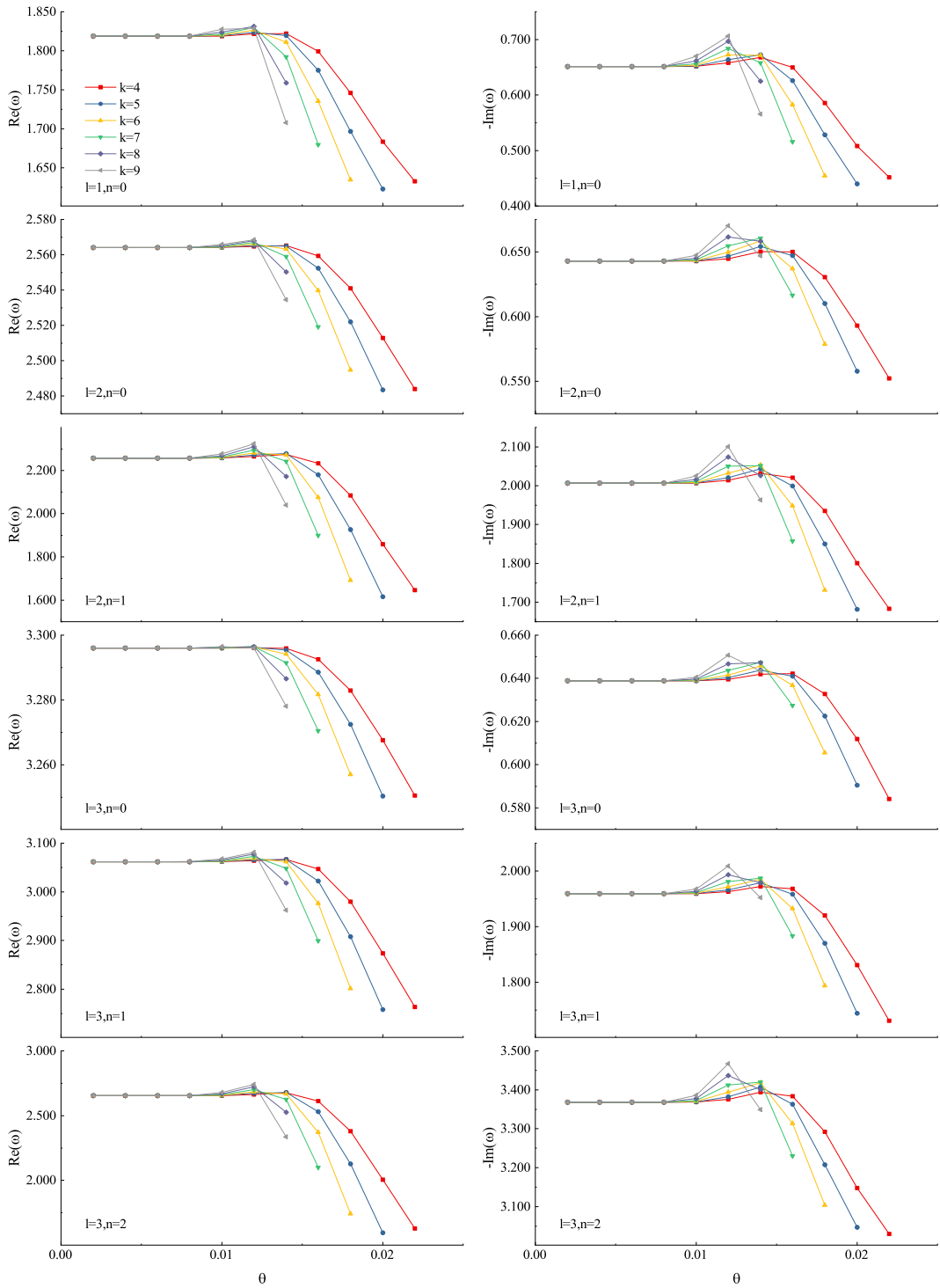}
\caption{Similar to the Fig. 4, but for  $D=6$ and $k=4, 5, 6, 7, 8, 9$.}
\end{figure}

\begin{figure}[htbp]
\centering
\includegraphics[height=20.5cm,width=15.6cm]{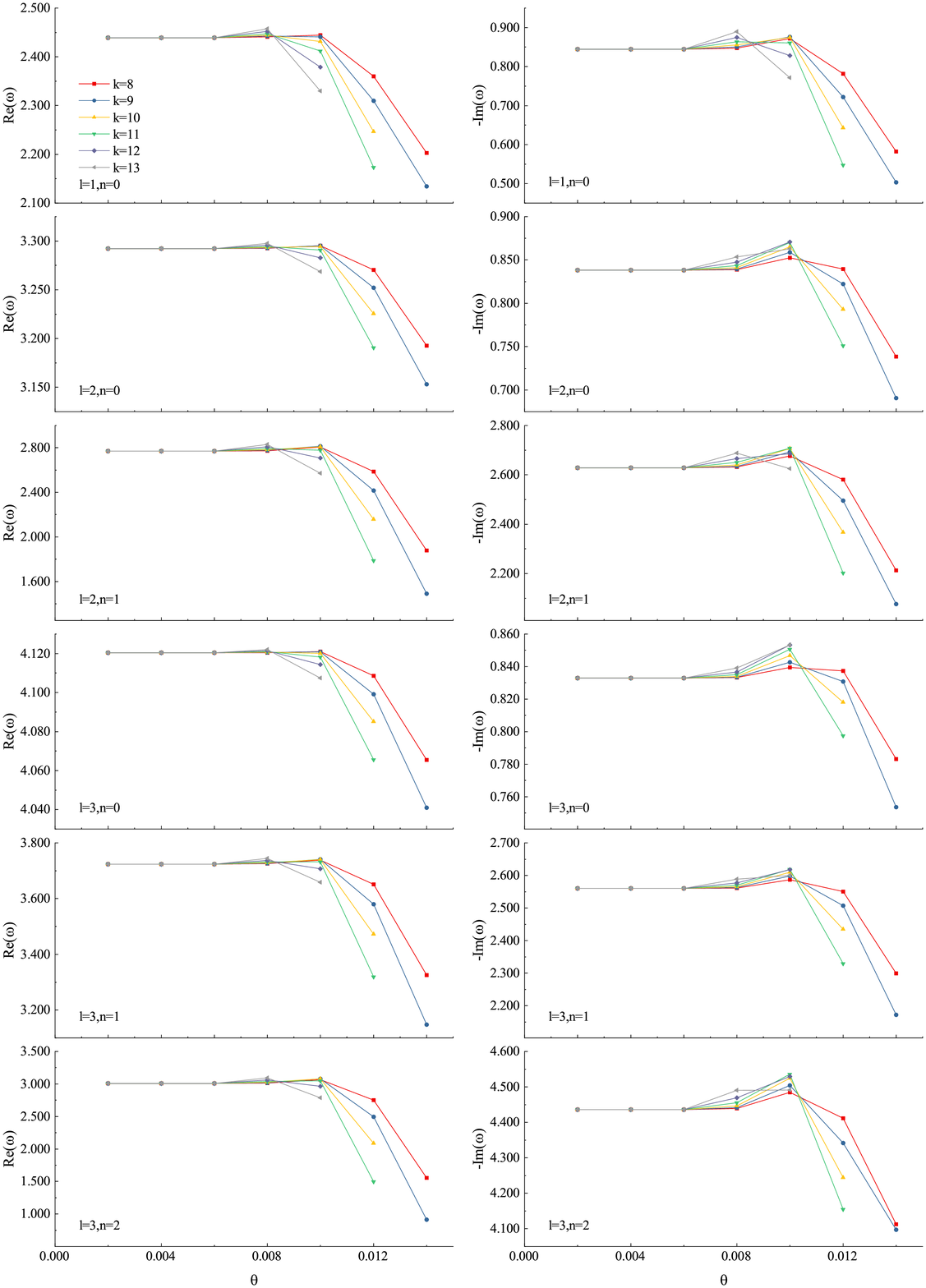}
\caption{Similar to the Fig. 4, but for  $D=7$ and $k=8, 9, 10, 11, 12, 13$.}
\end{figure}

\section{Evolution of perturbations analyzed with the time-domain integration method}

First of all, a brief introduction to the Gundlach-Price-Pullin method \cite{Gundlach:1993tp}, that is, the time-domain integration method or finite difference method.
In the time domain, we study the perturbation attenuation of noncommutative $D$-dimensional Schwarzschild black hole spacetime by using the numerical characteristic integral method, that
uses the light-cone variable $du=dt-dx$ and $dv=dt+dx$, and rewrite Eq. (\ref{waveequ}) as
\begin{equation}
-4 \frac{\partial^{2}}{\partial u \partial v} \Psi(u, v)=V_{i}\bigg[r\Big(\frac{v-u}{2}\Big)\bigg] \Psi(u, v).
\end{equation}
In the characteristic initial value problem, initial data are specified on the two null surfaces $u=u_{0}$ and $v=v_{0}$,
since the basic aspect of field attenuation has nothing to do with the initial conditions, it is assumed that the field $\Psi$ is initially in the form of Gaussian wave packets, so we choose the initial condition as $\Psi\left(u=u_{0}, v\right)=\exp \left[-\frac{\left(v-v_{c}\right)^{2}}{2 \sigma^{2}}\right]$, $\Psi\left(u, v=v_{0}\right)=0$, and choose the appropriate Gaussian wave package in the practical computation.
The discretization method we use is
\begin{equation}
\Psi(N)=\Psi(W)+\Psi(E)-\Psi(S)- \frac{\Delta^{2}}{8}\Big [\Psi(W)+\Psi(E) \Big]V(S)+\mathcal{O}\left(\Delta^{4}\right),
\end{equation}
where we have used the following definitions for the points: $N=(u+\Delta, v+\Delta)$, $W=(u+\Delta, v)$, $E=(u, v+\Delta)$ and $S=(u, v)$.
When the integration is completed, the value $\Psi\left(u_{\max }, v\right)$ is extracted, where $u_{\max }$ is the maximum value of $u$ on the numerical grid, as long as the $u_{\max }$ is large enough, we have a good approximation of the wave function at the event horizon.
In this way, we obtain the time-domain profile, which is a series of values of the perturbation field $\Psi(t=(v+u)/2, x=(v-u)/2)$ at a given position $x$ and discrete moments $t_{0}, t_{0}+h, t_{0}+2h, \cdots, t_{0}+Nh$.

In this paper, the time evolution behavior of scalar field perturbation of noncommutative $D$-dimensional Schwarzschild black holes was investigated.
In this paper, several smaller $k$ values are selected within the range of allowable $k$ corresponding to the dimension ($6 \geq D \geq 4$), and the effects of different parameters $k$ and noncommutative parameter $\theta$ on the time-domain profiles of the perturbation are compared.
It can be seen in Figs. 8-10 that $k$ and $\theta$ have little influence on it.
As shown in Fig. 11, the influence of different dimensions on the time-domain profiles of the perturbation is given, and it can be seen that the larger the dimension, the faster the decay.

\begin{figure}[htbp]
\centering
\includegraphics[height=6.2cm,width=16cm]{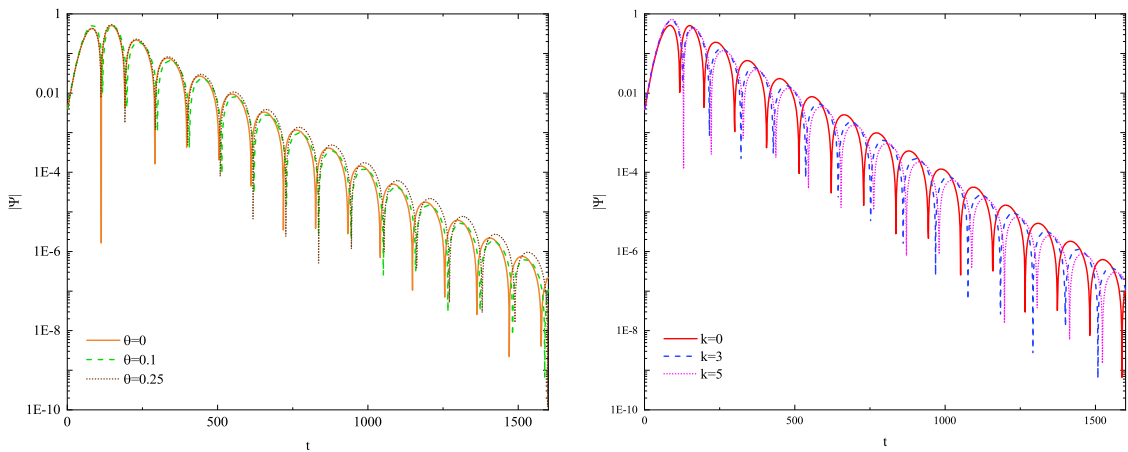}
\caption{Time-domain profiles of the massless scalar field perturbation corresponding to different $\theta$ when $k=0$ in the left panel. Time-domain profiles of the massless scalar field perturbation corresponding to different $k$ when $\theta=0.1$ in the right panel. The parameter is selected as $D=4$, $l=1$, $\sigma=3$, $v_{c}=10$.}
\end{figure}

\begin{figure}[htbp]
\centering
\includegraphics[height=6.2cm,width=16cm]{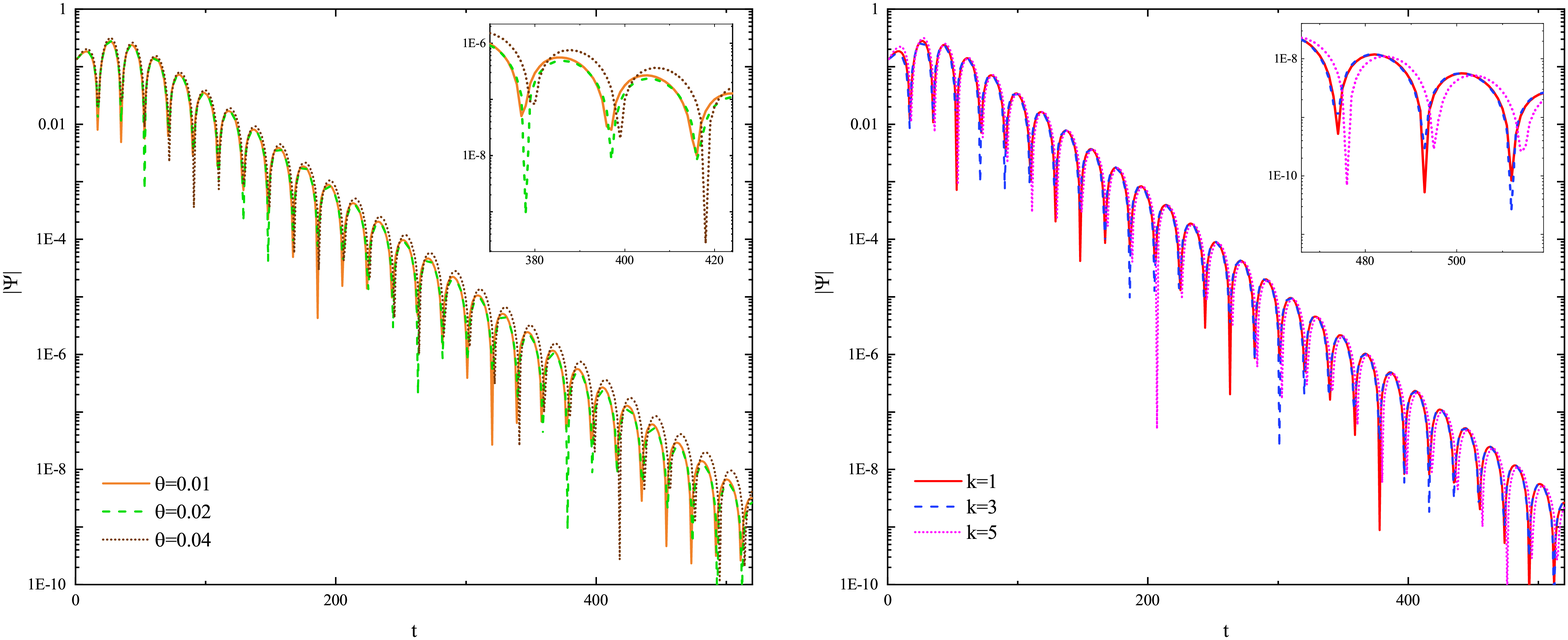}
\caption{Time-domain profiles of the massless scalar field perturbation corresponding to different $\theta$ when $k=1$ in the left panel. Time-domain profiles of the massless scalar field perturbation corresponding to different $k$ when $\theta=0.02$ in the right panel. The parameter is selected as $D=5$, $l=2$, $\sigma=1$, $v_{c}=2$.}
\end{figure}

\begin{figure}[htbp]
\centering
\includegraphics[height=6.2cm,width=16cm]{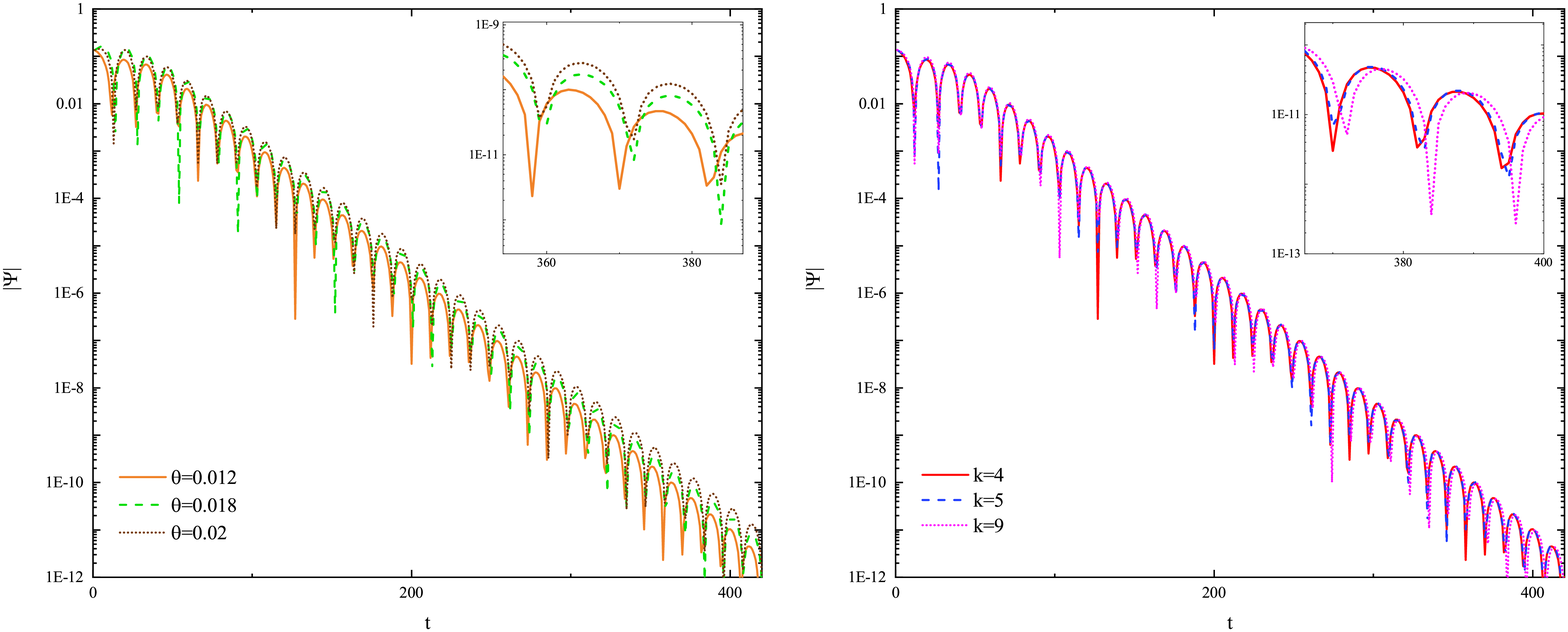}
\caption{Time-domain profiles of the massless scalar field perturbation corresponding to different $\theta$ when $k=4$ in the left panel. Time-domain profiles of the massless scalar field perturbation corresponding to different $k$ when $\theta=0.012$ in the right panel. The parameter is selected as $D=6$, $l=2$, $\sigma=1$, $v_{c}=2$.}
\end{figure}

\begin{figure}[htbp]
\centering
\includegraphics[height=9cm,width=13.5cm]{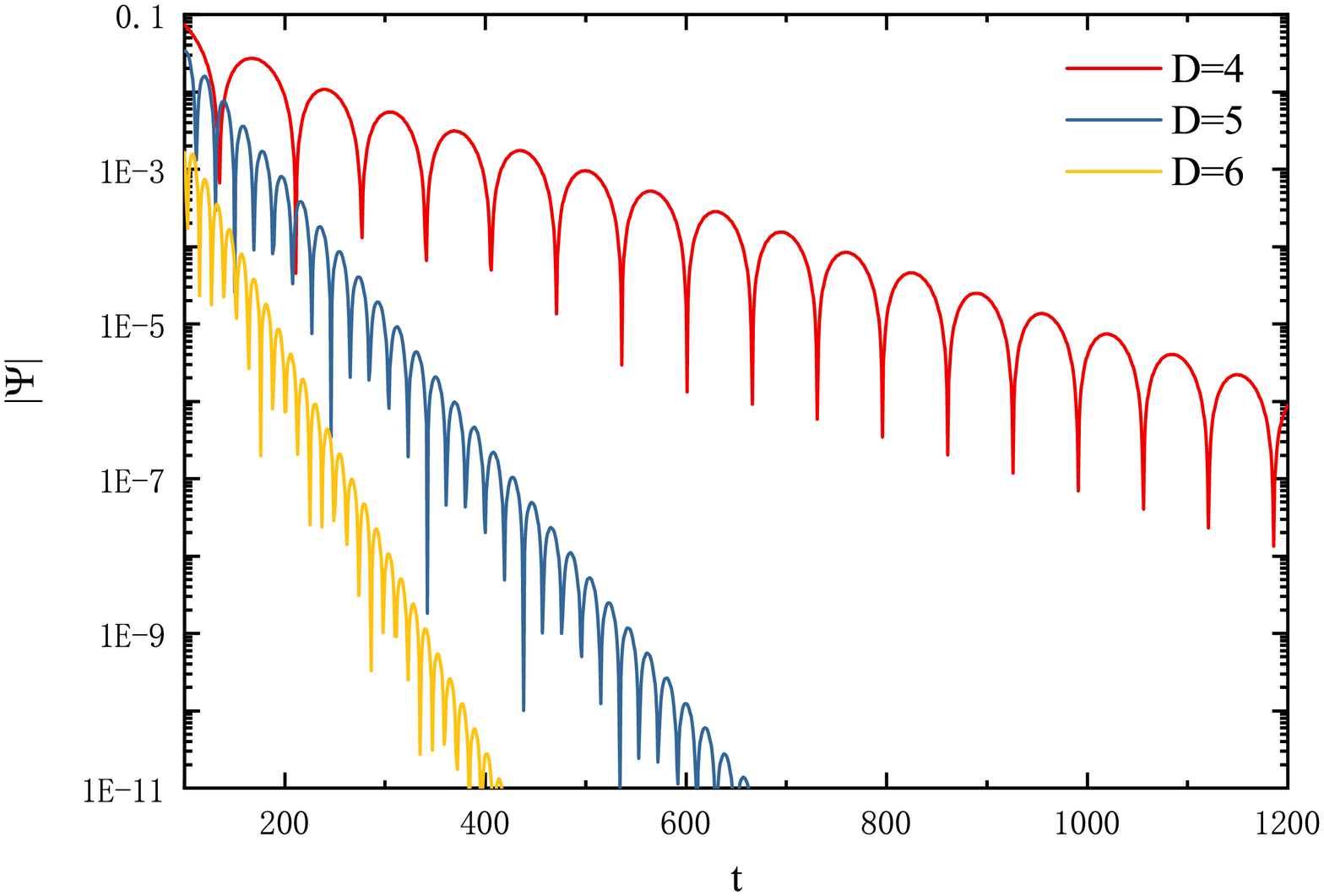}
\caption{Time-domain profiles of the massless scalar field perturbation corresponding to different $D$. The parameter is selected as $k=4$, $\theta=0.02$, $l=2$, $\sigma=1$, $v_{c}=2$.}
\end{figure}

\section{summary}
In this work, we have studied the quasinormal modes of noncommutative $D$-dimensional Schwarzschild-Tangherlini spacetime black hole with general smeared matter distribution in massless scalar field.
 Under the hoop conjecture, the allowable range of $k$ corresponding to different $D$ value is given. Then we find out the valid range of $\theta$ corresponds to different $D$ and $k$ values when the event boundary exists.  By means of three different numerical methods, we made a detailed analysis of the scalar QNM frequencies by varying the characteristic parameters of the scalar perturbation and the spacetime parameters of this black hole.
 Our results have shown that the QNMs obtained by the 3rd WKB method and the AIM method are  very close to each other, so we can reach a conclusion that the 3rd WKB approximation  is more accurate than other high order approximation under our consideration.

Next, the time evolution behavior of scalar field perturbation of noncommutative $D$-dimensional Schwarzschild black holes is studied.
Numerical results show the influence of different spacetime parameters $\theta$ and $k$ on the time-domain profiles of the perturbation. It is found that the influence of the parameters $\theta$ and  $k$ on the QNMs is small and negligible.

Finally, we systematically elaborate on the influence of $D$, $\theta$ and $k$ on QNMs.
$Re(\omega)$ and $-Im(\omega)$ increase significantly with the increase of $D$, as shown in Table. \ref{DSchdata} and Fig. 11.
As $\theta$ increases, the values of $Re(\omega)$ and $-Im(\omega)$ increase slightly and then decrease, and both decrease significantly when the $\theta$ approaches the extreme $\theta_{\max}$. as shown in Fig. 2-10.
When $\theta$ is small, $Re(\omega)$ and $-Im(\omega)$ remain unchanged with the change of $k$.
When the $\theta$ is large, although there is such a situation that $Re(\omega)$ and $-Im(\omega)$ increase with the increase of $k$, it mainly shows a decreasing trend, and this change is not significant. as shown in Fig. 8-10.

\begin{acknowledgments}
The authors thank Dr. X. Zhang and C. Feng for their positive help and useful discussion. We like to thank R. A. Konoplya for providing the WKB approximation. We also like to thank developers of the AIM method for their opening codes.  This work was supported by the National Key Research and Develop Program of China under Contract No. 2018YFA0404404 and  the Key  Research Program of the Chinese Academy of Sciences (Grant No. XDPB09-02).
\end{acknowledgments}

\bibliography{NDSch-Ref}% Produces the bibliography via BibTeX.

\end{document}